\RequirePackage{fix-cm}
\documentclass[smallextended]{svjour3}       % onecolumn (second format)
\smartqed  % flush right qed marks, e.g. at end of proof
\usepackage{graphicx}

\usepackage{etex}
\reserveinserts{18}
\usepackage{morefloats}
\usepackage{amssymb}
\usepackage{graphicx}
\usepackage{amsmath}
\usepackage{collectbox}

\usepackage{url}
\usepackage[pdfpagelabels,hypertexnames=false,breaklinks=true,bookmarksopen=true,bookmarksopenlevel=2]{hyperref}
\usepackage{float}
\usepackage{graphics}
\usepackage{enumerate}
\usepackage{multicol}
\usepackage{csquotes}
\usepackage{subcaption}
\usepackage{booktabs}
\usepackage{array}
\usepackage{fixltx2e}
\usepackage[usenames]{color}
\usepackage{listings}
\usepackage{lineno}
\newcommand{\RQOne}{Do developers change behaviour in the context of impolite/negative comments?}
\newcommand{\RQTwo}{What is the probability of shifting from
comments holding positive emotions to comments holding negative emotion?}

\begin{document}

\title{Analysing Developers Affectiveness through Markov chain Models}


\author{Giuseppe Destefanis\\
        Marco Ortu,\\
        Steve Counsell,\\
        Stephen Swift, \\
        Roberto Tonelli, \\
        Michele Marchesi \\
}

\authorrunning{Destefanis et al.}

\institute{Giuseppe Destefanis \at
              University of Hertfordshire\\
              \email{g.destefanis@herts.ac.uk}    
           \and
           Steve Counsell, Stephen Swift \at
              Brunel University London\\
              \email{\{steve.counsell,stephen.swift\}@brunel.ac.uk} 
           \and
           Marco Ortu, Roberto Tonelli, Michele Marchesi \at
              DIEE, University of Cagliari\\
              \email{\{marco.ortu,roberto.tonelli,michele\}@diee.unica.it}}


\maketitle

\begin{abstract}
In this paper, we present an analysis of more than 500K comments from open-source
repositories of software systems.
Our aim is to empirically determine how developers interact with each other
under certain psychological conditions generated by politeness, sentiment and
emotion expressed within developers' comments.
Developers involved in an open-source projects do not usually know each other; they mainly communicate through mailing lists, chat rooms, 
and tools such as issue tracking systems.
The way in which they communicate affects the development process and the productivity of the people involved in the project.
We evaluated politeness, sentiment and emotions of comments posted by developers and studied the communication flow to understand how 
they interacted in the presence of impolite and negative comments (and \textit{vice versa}).
Our analysis shows that when in presence of impolite or negative comments, the probability of the next comment 
being impolite or negative is 14\% and 25\%, respectively;  \textit{anger} however,
has a probability of 40\% of being followed by a further \textit{anger} comment.
The result could help managers take control the development phases of a system, since social aspects can seriously affect a developer's productivity. In a distributed environment this may have a particular resonance.
\end{abstract}

\section{Introduction}

The study of emotions and psychological status of developers and people involved in the software-building system is gaining the attention of both practitioners 
and researchers \cite{ke2010effects}. 
Feldt et al. \cite{feldt2008towards} focused on personality as one important psychometric factor and presented initial results from an empirical study investigating 
the correlation between personality and attitudes to software engineering processes and tools.
 
Software is a complex artefact which requires sharing of knowledge, team building and exchange of opinion between people. 
While it has been possible to standardise classical industrial processes (e.g., car production), it is still difficult to standardise software production. 
Immateriality plays a major role in the complexity of software and despite attempts to standardise the software production process, software engineering 
is still a challenging and open field.
There are too many constraints to take into account.
Developers build an artefact that will be executed on a machine; software metrics, design patterns, micro patterns and good practices help to increase the quality of a software \cite{concas2013micro,destefanis2012analysis,destefanis2012micro}, but developers are humans and prone to human sensitivities. 
Coordinating and structuring developer teams is a vital activity for software companies \cite{ortujira} and dynamics within a team have a direct influence 
on group success; on the other hand, social aspects are intangible elements which, if monitored, can help the team in reaching its goals.
Researchers are increasingly focusing their effort on understanding how the human aspects of a technical discipline can affect the final results 
\cite{brief2002organizational,erez2002influence,graziotin2015understanding,kaluzniacky2004managing}.

Open-source development usually involves developers that voluntarily participate in a project by contributing with code. The management of such developers 
could even be more complex than the management of a team within a company, since developers are not in the same place at the same time and coordination 
becomes more difficult.
The absence of face-to-face communication mandates the use of mailing lists, electronic boards, or specific tools such as issue tracking systems. Being rude when 
writing a comment or replying to a contributor can affect the cohesion of the group and the successfulness of a project; equally a respectful environment is an incentive 
for new contributors joining the project \cite{rigby2007can,tourani2014monitoring}.

In this paper, we empirically analyze more than 500K comments from Ortu et al. \cite{ortu2015jiradataset,ortu2016jiraMSR} to understand how agile developers behave when dealing with polite/impolite or 
positive/negative (sentiment) issue comments. 
We empirically built three Markov chain models with states for politeness (polite, neutral, impolite), sentiment (positive, neutral, negative), and emotions (joy, anger, love, sadness) for each projects in our corpus and three general Markov chain which generalize our findings.
We aim to answer the following questions: \\

\textbf{RQ1:} \RQOne \\

Developers tended to answer to impolite comments with a polite comment with higher probability than impolite comments. Similar result applies for sentiment, developers tended to answer to negative comments (negative in term of sentiment) with a positive comment with higher probability than negative.\\

\textbf{RQ2:} \RQTwo \\

Negative emotions such as \textit{sadness} and
\textit{anger} tend to be followed by negative emotions more than positive
emotion are followed by positive emotions.\\

This paper is an extended version of our previous paper \cite{ortu2016arsonists} accepted for presentation at the 17th International Conference on Agile Software Development (XP 2016) in Edinburgh, Scotland, May 24-27, 2016. Compared with our XP 2016 paper, this version extends the study for every single project present in the considered dataset. We built 45 transition matrices (three for each system in the corpus) and explained how we have obtained the three final general Markov chains for Politeness, Sentiment and Emotion.

The remainder of this paper is structured as follows: in the next section, we provide a summary of related work. Section 3 describes the dataset used for this study and our approach/rationale to evaluate affectiveness of comments posted by developers. In section 4, we present the methodology used for building the Markov chains. In Section 5, we present the results and elaborate on the research questions we address in Section 6. Section 7 discusses the threats to validity. We summarize the study findings in Section 8 and present detailed results about the 45 Markov chains in the appendix section.

\section{Related Work}

Several recent studies have demonstrated the importance and relationship of productivity and quality to human aspects associated with the software development process.
Ortu et al. studied the effect of politeness \cite{destefanis2016software,ortu2015would} and emotions \cite{ortubullies} on the time required to fix any given issue. The authors 
demonstrated that emotions did have an effect on the issue fixing time. Namely, results showed that issue fixing time for polite issues was shorter than issue fixing time for impolite and mixed-issues (polite and impolite issues). If someone is asked to accomplish a task in a polite way, there is higher possibility for a relaxed collaboration and faster results. On the other hand, impolite requests can easily generate discomfort, stress and burnout, often negatively impacting the actual time taken to complete a given task. The study also showed that a positive correlation existed between the percentage of polite comments and Magnetism and Stickiness of a project (higher attractiveness) and that the percentage of polite comments over time was (for the majority of the projects in our corpus) seasonal and not random.

Research has focused on understanding how the human aspects of a technical discipline can affect final results \cite{brief2002organizational,erez2002influence,ortu2017diverse,kaluzniacky2004managing,rousinopoulos2014sentiment}, and the effect of politeness \cite{novielli2014towards,tan2014rude,tsay2014let}.
The Manifesto for Agile Development indicates that people and communications are more essential than procedures and tools \cite{beck2001manifesto}.
Several recent studies have demonstrated the importance and relationship of productivity and quality to human aspects associated with the software development process.
Ortu et al. studied the effect of politeness \cite{ortu2015would} and emotions \cite{ortubullies} on the time required to fix any given issue. The authors 
demonstrated that emotions did have an effect on the issue fixing time.
Steinmacher et al. \cite{steinmacher2015social} analyzed social barriers that obstructed first contributions of newcomers (new developers joining an open-source project). The study indicated how impolite answers were considered as a barrier by newcomers. 
These barriers were identified through a systematic literature review, responses collected from open source project contributors and students contributing to open source projects.
Rigby et al. \cite{rigby2007can} analyzed, using a psychometrically-based linguistic analysis tool, the five big personality traits of software developers in the Apache httpd server mailing list.
The authors found that the two developers that were responsible for the major Apache releases had similar personalities and their personalities were different from other developers. Tourani et al.~\cite{Tourani2014} evaluated the usage of automatic sentiment analysis to identify distress or happiness in a development team.  The authors mined sentiment values from the mailing lists of two mature projects of the Apache software foundation considering both users and developers. The results showed that sentiment analysis tools obtained low precision on emails written by developers due to ambiguities in technical terms and difficulties in distinguishing positive or negative sentences from neutral.\\ 
M{\"a}ntyl{\"a} et al. \cite{mika} explored the Valence - Arousal - Domincance (VAD) metrics and their properties on 700,000 Jira issue reports containing over 2,000,000 comments. Using a general-purpose lexicon of 14,000 English words with known VAD scores, the results showed that issue reports of different type (e.g., Feature Request vs. Bug) had a fair variation of Valence, while increase in issue priority (e.g., from Minor to Critical) typically increased Arousal. Furthermore, the results showed that as an issue’s resolution time increased, so did the arousal of the individual the issue was assigned to. Finally, the resolution of an issue increased valence, especially for the issue Reporter and for quickly addressed issues.\\ 
Bazzelli et al. \cite{bazelli2013personality} analyzed questions and answers on stackoverflow.com to determine the developer personality traits, using the Linguistic Inquiry and Word Count \cite{pennebaker2001linguistic}.
The authors found that the top reputed authors were
more extroverted and expressed less negative emotions than authors of down voted posts.
Jurado and Rodriguez~\cite{Jurado2015} gathered the issues of nine high profile software projects hosted on GitHub. Through an analysis of the occurrence of Ekman's~\cite{Ekman1992} basic emotions among the projects and issues, the authors discovered that in open source projects, sentiments expressed in the form of joy are % pervading and are present at 
almost one magnitude of order more common than the other basic emotions. Still, more than 80\% of the content was not classified as exhibiting a high amount of sentiment. Several other studies have been conducted using sentiment analysis and emotion mining for analyzing app reviews, in order to gain insights such as ideas for improvements, user requirements and to analyze customer satisfaction (e.g.,~\cite{guzman2014users,maalej2015bug}).

Guzman et al. \cite{guzman2014sentiment} performed sentiment analysis of Github’s commit comments to investigate how emotions were related to a project’s programming language, the commits’ day of the week and time, and the approval of the projects. The analysis was performed over 29 top-starred Github repositories implemented in 14 different programming languages. The results showed Java to be the programming language most associated with negative affect. No correlation was found between the number of Github stars and the affect of the commit messages. Garcia et al. \cite{garcia2013role} analyzed the relation between the emotions and the activity of contributors in the Open Source Software project Gentoo. The case study built on extensive data sets from the project’s bug tracking platform Bugzilla, to quantify the activity of contributors, and its mail archives, to quantify the emotions of contributors by means of sentiment analysis. The Gentoo project is known for a period of centralization within its bug triaging community. This was followed by considerable changes in community organization and performance after the sudden retirement of the central contributor. The authors analyzed how this event correlated with the negative emotions, both in bilateral email discussions with the central contributor, and at the level of the whole community of contributors. The authors also extended the study to consider the activity patterns of Gentoo contributors in general. They found that contributors were more likely to become inactive when they expressed strong positive or negative emotions in the bug tracker, or when they deviated from the expected value of emotions in the mailing list. The authors used these insights to develop a Bayesian classifier which detected the risk of contributors leaving the project.

Pletea et al. \cite{pletea2014security} studied security-related discussions on GitHub, as mined from discussions around commits and pull requests. The authors found that security-related discussions account for approximately 10\% of all discussions on GitHub and that more negative emotions were expressed in security-related discussions than in other discussions. These findings confirmed the importance of properly training developers to address security concerns in their applications as well as the need to test applications thoroughly for security vulnerabilities in order to reduce frustration and improve overall project atmosphere.

Panichella et al. \cite{panichella2015can} presented a taxonomy to classify app reviews into categories relevant to software maintenance and evolution, as well as an approach that merges three techniques (Natural Language Processing, Text Analysis, Sentiment Analysis) to automatically classify app reviews into the proposed categories. The authors showed that the combined use of these techniques achieves better results (a precision of 75\% and a recall of 74\%) than results obtained using each technique individually (precision of 70\% and a recall of 67\%).

Gomez et al. \cite{gomez2012does} performed an experiment to evaluate whether the level of extraversion in a team influenced the final quality of the software products obtained and the satisfaction perceived while this work was being carried out. Results indicated that when forming work teams, project managers should carry out a personality test in order to balance the amount of extroverted team members with those who are not extraverted. 
This would permit the team members to feel satisfied with the work carried out by the team without reducing the quality of the software products developed.

Hidden Markov Models (HMMs) have been largely used in application for speech recognition, handwriting recognition, bioinformatic, textures and patterns detection. Markov Chains belong to the category of Observable Markov models and this means that the states are directly visibile by an external observer and that it it possible to relate a physical event to each state.
HMMs extends these models for cases in which the observation is a probabilistic function of a state. 
It is not possible to see the physical event responsible for generating an observation, but it is possible to observe only the result of the event. The model is not an observable stochastic process, and it is possible to model it through a set of stochastic processes which produce the observations sequence.

Novielli \cite{novielli2010hmm} used Hidden Markov Models as a formalism to represent differences in the dialogue model among different categories of users or engagement. The author proposed a corpus-based approach to train HMMs from natural dialogues. 
The models were validated through a leave-one-out testing procedure and real-time use of these models were implemented using a stepwise approach. 
 The study investigated whether and how it was possible to recognise the user’s level of engagement by modeling the impact of the user’s attitude on the overall dialogue pattern. The paper proposed an application of the described approach in the advice-giving domain.  Results presented HMMs as a promising and powerful formalism for representing differences in the structure of the interaction with subjects experiencing different levels of engagement. The main differences were observed during the persuasion phase, in which users clearly differentiate their behavior according to their engagement in the advice-giving task. 
 
Wu et al. \cite{wu1998spoken}, proposed a corpus-based HMMs to model the intention of a sentence and evaluated it investigating a spoken dialogue model for air travel information service. Each intention was represented by a sequence of word segment categories determined by a task-specific lexicon and a corpus.
Five intention HMMs were defined during the training procedure. In the intention identification process, the phrase sequence was fed to each HMMs intention. Given a speech utterance, the Viterbi algorithm was used to find the most likely intention sequences. The intention HMM considered the phrase frequency and the syntactic and semantic structure in a phrase sequence. 
The experiments were carried out using a test database from 25 speakers (15 male and 10 female), with 120 dialogues containing 725 sentences in the test database. The experimental results showed that the correct response rate can achieve about 80.3\% using intention HMMs.

Compared to the existing literature, the goal of this paper is to build Markov chain models which describe how developers interact in a distributed environment evaluating politeness, sentiment and emotions. Such models provide a mathematical view of the behavioural aspects among developers.

\section{Experimental Setup}
\label{sec:experimental_setup}
\subsection{Dataset}
\label{sec:dataset}

We built our dataset from fifteen open-source, publicly available projects from a dataset proposed by Ortu et al. \cite{ortu2015jiradataset,ortu2016jiraMSR} extracted from the Jira ITS of four popular open source ecosystems (as well as the tools and infrastructure used for extraction), i.e., the Apache Software Foundation, Spring, JBoss and CodeHaus communities. 

The dataset hosts more than 1K projects, containing more than 700K issue reports and more than 2 million issue comments. With such a wide dataset, the authors found that comments posted by developers contain not only technical information, but also valuable information about sentiments and emotions. 
Issue tracking systems store valuable data for testing hypotheses concerning maintenance, building statistical pre- diction models and the social interactions of developers when interacting with peers. In particular, the Jira Issue Tracking System (ITS) is a proprietary tracking system that has gained a tremendous popularity in the last years and offers unique features like a project management system and the Jira agile kanban board.

The dataset by Ortu et al. \cite{ortu2015jiradataset} contains the following tables:

\begin{itemize}
\item \textbf{ISSUES\_REPORT}. It stores the information extracted from the
issue reports. Issues are associated with comments and attachments and history
changes.
\item \textbf{ISSUES\_COMMENTS}. It represents all the comments posted by users in a Jira issue report. This table is associated with the
\textit{ISSUES\_REPORT} table.
\item \textbf{ISSUE\_BOT\_COMMENT}. It represents automatically generated
comments from tools such as Jenkins or Jira itself.
\item \textbf{ISSUES\_FIXED\_VERSION}. It records the software version of fixed
issues.
\item \textbf{ISSUES\_AFFECTED\_VERSION}. It records the software version affected by issues.
\item \textbf{ISSUE\_ATTACHMENT}. It represents all files attached to an issue
report.
\item \textbf{ISSUE\_CHANGELOG\_ITEM}. It represents all operations made on an
issue such as editing, updating, status changing, etc.
\end{itemize}

We selected the fifteen projects with the highest number of
comments (from December 2002 to December 2013), from those
projects which had a significant amount of activities in their agile kanban-boards. The projects were developed following agile practices (mainly continuous delivery and use of kanban-boards).
Table \ref{tab:datasetStatistics} shows summary project statistics.

\begin{table}[htp!] 
\centering  
\setlength{\tabcolsep}{14pt}
\begin{tabular}{lcc}    \toprule
\textbf{Project} & \textbf{\# of comments} & \textbf{\# of developers}  \\\midrule
HBase  				&  	91016  &  	951 \\
Hadoop Common  		&  	61958  &  	1243 \\
Derby  				&  	52668  &  	675 \\
Lucene Core  		&  	50152  &  	1107 \\
Hadoop HDFS  		&  	42208  &  	757 \\
Cassandra  			&  	41966  &  	1177 \\
Solr  				&  	41695  &  	1590 \\
Hive  				&  	39002  &  	850 \\
Hadoop Map/Reduce  	&  	34793  &  	875 \\
Harmony  			&  	28619  &  	316 \\
OFBiz  				&  	25694  &  	578 \\
Infrastructure   	&  	25439  &  	1362 \\
Camel  				&  	24109  &  	908 \\
Wicket				&	17449	&  1243 \\
ZooKeeper			&   16672  &    495 \\\bottomrule
 \hline
\end{tabular}  
\caption{Selected Project Statistics}
\label{tab:datasetStatistics}
\end{table}  

\subsection{Affective Metrics}
%\vspace{-0.5cm}
\label{sec:affectiveMetrics}

Henceforward, we consider the term ``affective metric'' as a definition indicating all
those measures linked to human aspects and obtained from text written by developers (i.e., comments posted on issue tracking systems).
This study is based on the affective metrics sentiment, politeness and emotions used by Ortu et al. \cite{ortubullies} and Destefanis et al. \cite{destefanisrandomness}, which have been shown being not correlated. 

\subsubsection{Sentiment.}
\label{sec:exp_sentiment}

We measured sentiment using the SentiStrength\footnote{http://sentistrength.wlv.ac.uk} tool, which is able to estimate the degree of positive and negative sentiment 
in short texts, even for informal language. SentiStrength, by default, detects two sentiment polarizations:
\begin{itemize}
\item Negative: -1 (slightly negative) to -5 (extremely negative)
\item Positive: 1 (slightly positive) to 5 (extremely positive)
\end{itemize}

The tool uses a lexicon approach based on a list of words to detect sentiment; SentiStrength was originally developed for the English language and was optimized 
for short social web texts. 
We used the tool to measure the sentiment of developers in issue comments.

\subsubsection{Politeness.}
\label{sec:exp_politeness}

To evaluate the level of politeness of comments related to a given issue, we used the tool developed by Danescu et al. \cite{Danescu-Niculescu-Mizil+al:13b}; the tool uses a machine learning approach and calculates the politeness of sentences providing, as a result, one of two possible labels: polite or impolite. 
The tool provides a level of confidence related to the probability of a politeness class being assigned. 
To prepare the training set for the machine learning approach, over 10,000 utterances were labeled using Amazon Mechanical Turk. They decided to restrict the residence of the annotators to the U.S. and conducted a linguistic background questionnaire. However, the annotators analysed comments written by authors from around the world and not only from the U.S. Therefore, the possible bias introduced by annotators with a similar cultural background is reduced and the different cultures of the developers involved in the analysis are considered. The use of the tool would have been problematic, if annotators were from the U.S. and they had analysed only comments written by authors from the U.S. Danescu et al. \cite{Danescu-Niculescu-Mizil+al:13b} evaluated the classifiers both in an in-domain setting, with a standard leave-one-out cross validation procedure, and in a cross-domain setting, where they trained on one domain and tested on the other.\\
They have compared two classifiers, a bag of words classifier (BOW) and a linguistic informed classifier (Ling.) and used a human labellers as a reference point.
Table \ref{tab:accuracies} (from Danescu et al. \cite{Danescu-Niculescu-Mizil+al:13b}) shows the accuracies of the two classifiers for Wikipedia and Stack Exchange, for in-domain and cross-domain settings. Human performance is included as a reference point.

\begin{table}[htp!] 
\centering  
\setlength{\tabcolsep}{14pt}
\begin{tabular}{ccccc}    \toprule
& \multicolumn{2}{c}{In-domain} & \multicolumn{2}{c}{Cross-domain} \\
\midrule
Train  	& Wiki & SE  & Wiki & SE  \\
Test  	& Wiki & SE & SE & Wiki \\\hline
BOW  	& 79.84\% & 74.47\% & 64.23\% & 72.17\% \\
Ling. 	& 83.79\% &  78.19\% & 67.53\% & 75.43\%\\\hline
Human	& 86.72\% & 80.89\% & 80.89\% & 86.72\% \\\bottomrule
 \hline
\end{tabular}  
\caption{Accuracies of the two classifiers for Wikipedia and Stack Exchange}
\label{tab:accuracies}
\end{table}

We considered comments whose level of confidence was less than 0.5 as neutral (the text did not convey either politeness or impoliteness).
For each comment we assigned a value according to the following rules:

\begin{itemize}
\item Value of +1 for comments marked as polite; 
\item Value of 0 for comments marked as neutral (confidence level$<$0.5);
\item Value of -1 for comments marked as impolite.
\end{itemize}

For each issue in our dataset, we built a temporal series of comments, and using the two tools we assigned a value of politeness and sentiment for each comment 
in the series.
Next, for each issue, we calculated, starting from the first comment posted, the probability of having a polite/impolite/neutral following comment (for politeness), 
and a positive/neutral/negative comment (for sentiment).
We thus calculated the probability of shifting from ``polite'' to ``neutral'' and \textit{vice versa}; from ``polite'' to ``impolite'' and \textit{vice versa}; finally, 
from ``neutral'' to ``impolite'' and \textit{vice versa}.

\subsubsection{Emotion.}
\label{sec:exp_emotion}

The presence of emotion in software engineering artifacts have been analysed by Destefanis et al. \cite{destefanis2016software} and Ortu et al. \cite{ortu2015would}.

Ortu et al. \cite{ortubullies}
provided a machine learning based approach for emotion detection in developers'
comments based on Parrott’s emotional framework, which consists of six basic emotions: joy, sadness, love, anger, sadness, and fear. For each of the four emotions, the authors built a dedicated Support Vector Machine classifier and used a manually annotated corpus of comments and their emotion for training the machine learning Classifiers, one for each emotion. The training set consisted of 4000 sentences (1000 for each emotion), manually annotated by three raters having a strong background in computer science. Elfenbein et al. \cite{elfenbein2002universality} provided evidence that for members of the same cultural and social group it is easier to recognise emotions than for people belonging to different groups.

We used the emotion detection tool provided by Ortu et al. \cite{ortubullies} to detect the presence of \textit{sadness}, \textit{anger},
\textit{joy}, \textit{love} and \textit{neutral}.
 
Table \ref{tab:emotions_example} shows several examples of emotions detected from comments in our dataset. 

\begin{table}[htp!]
\centering
\begin{tabular}{|m{10cm}|c|}    \hline
Comments        		&  Emotion \\\hline
\begin{lstlisting}
1. Thanks for your input! You’re, like, awesome
2. Thanks very much! I appreciate your efforts
3. I would love to try out a patch for [... ] \end{lstlisting} &  	  	\textbf{Love} \\\hline
\begin{lstlisting}
1. I’m happy with the approach and the code looks good
2. great work you guys!
3. Hope this will help in identifying more usecases\end{lstlisting} &  	  	\textbf{Joy} \\\hline
\begin{lstlisting}
1. I will come over to your work and slap you
2. WT*, a package refactoring and class renaming in a patch?
3. This is an - ugly - workaround
\end{lstlisting} &  	\textbf{Anger} \\\hline
\begin{lstlisting}
1. Sorry for the delay Stephen.
2. Sorry of course printStackTrace() wont work
3. wish i had pay more attention in my english class .... now its pay back time .... :-(
4. Apache Harmony is no longer releasing. No need to fix this, as sad as it is.
\end{lstlisting} &  	  	\textbf{Sadness} \\\hline

\end{tabular}
\caption{Examples of comments expressing emotions.}
\label{tab:emotions_example}
\end{table}

Craggs et al. \cite{craggs2004two} argued that it is necessary to select a correct number of unit types into which a dialogue can be segmented and to which annotation can be applied, often the choice of unit is obvious, but this is not the case for emotion, since \textit{``emotional episodes exist over an indistinct period of time, fading in and out and subtly changing throughout a dialogue.''} and proposed to distinguish between the intensity of the emotion and its polarity using a two dimensional annotation scheme for emotion in dialogue. The author claim that the scheme is equally applicable to dialogues conducted in different domains and languages, but there is no evidence which supports this claim in the software engineering context.
In contrast to Craggs et al. \cite{craggs2004two}, Murgia et al. \cite{Murgia:2014} found that providing human raters with more context about an issue seems to cause doubt (i.e., nuances) instead of more confidence in the identified emotions.

\section{Affective Markov Chains}

\label{sec:exp_markov_chains}

Markov Chains (MC) have been used to model behavioural aspects in social sciences \cite{jordan1998learning}, \cite{snijders2001statistical}. 
Markov Chains  provide a unique representation where the transitions between the different states of Sentiment, Politeness and Emotion, extracted from developers comments, can be modelled; it implicitly assumes ``memorylessness'' (through the Markov property) and this  simplifies our analysis. Furthermore the models allow a straightforward determination of probability transitions between emotional states; this is crucial in determining how developer behaviour is influenced by other developer moods

A Markov chain consists of \textit{X} states and is a discrete-time stochastic process, a process that occurs in a series of time-steps in each of which a random choice is made.
A Markov chain can be represented with graphs and/or matrices. The states can be represented with circles (nodes or vertices) and directed edges (links) 
connecting node $i$ and node $j$ if $p_{ij}>0$.

From each node (or vertice) there is at least an out-coming edge: some nodes have links connecting to themselves, some nodes cannot be connected with each other, 
while some can be connected through bidirectional links.   
A probability $p_{ij}$ is associated to each connection $i\rightarrow j$, defined as the transition probability from state $i$ to $j$.
The following properties need to be satisfied for $p_{ij}$ values:

\[
p_{ij} \in [0,1], \qquad \sum_{j} p_{ij} = 1 \enspace \forall i
\]

A (square) transition matrix \textbf{P} contains only positive elements, with sum equal to 1 for each row (1). 

 \[\mathbf{P} =\begin{bmatrix}
  p_{11} &p_{12}&\cdots &p_{1n} \\
  p_{21}&p_{22}&\cdots &p_{2n} \\
  \vdots & \vdots & \ddots & \vdots\\
  p_{n1}&p_{n2}&\cdots &p_{nn}
  \end{bmatrix}\]\\

$p_{11}$ indicates the probability of staying in the state 1, $p_{12}$ indicates the probability of moving from state 1 to state 2 and $p_{1n}$ indicates the probability of moving from state 1 to state n.

\begin{eqnarray}
p_{11}+p_{12}+\cdots+p_{1n}=1 \notag\\
p_{21}+ p_{22}+ \cdots= + p_{2n}=1 \\
p_{n1}+p_{n2}+\cdots+p_{nn}=1 \notag
\end{eqnarray}
\\

\begin{figure}[htp!]
\centering
\includegraphics[width=12cm,height=!]{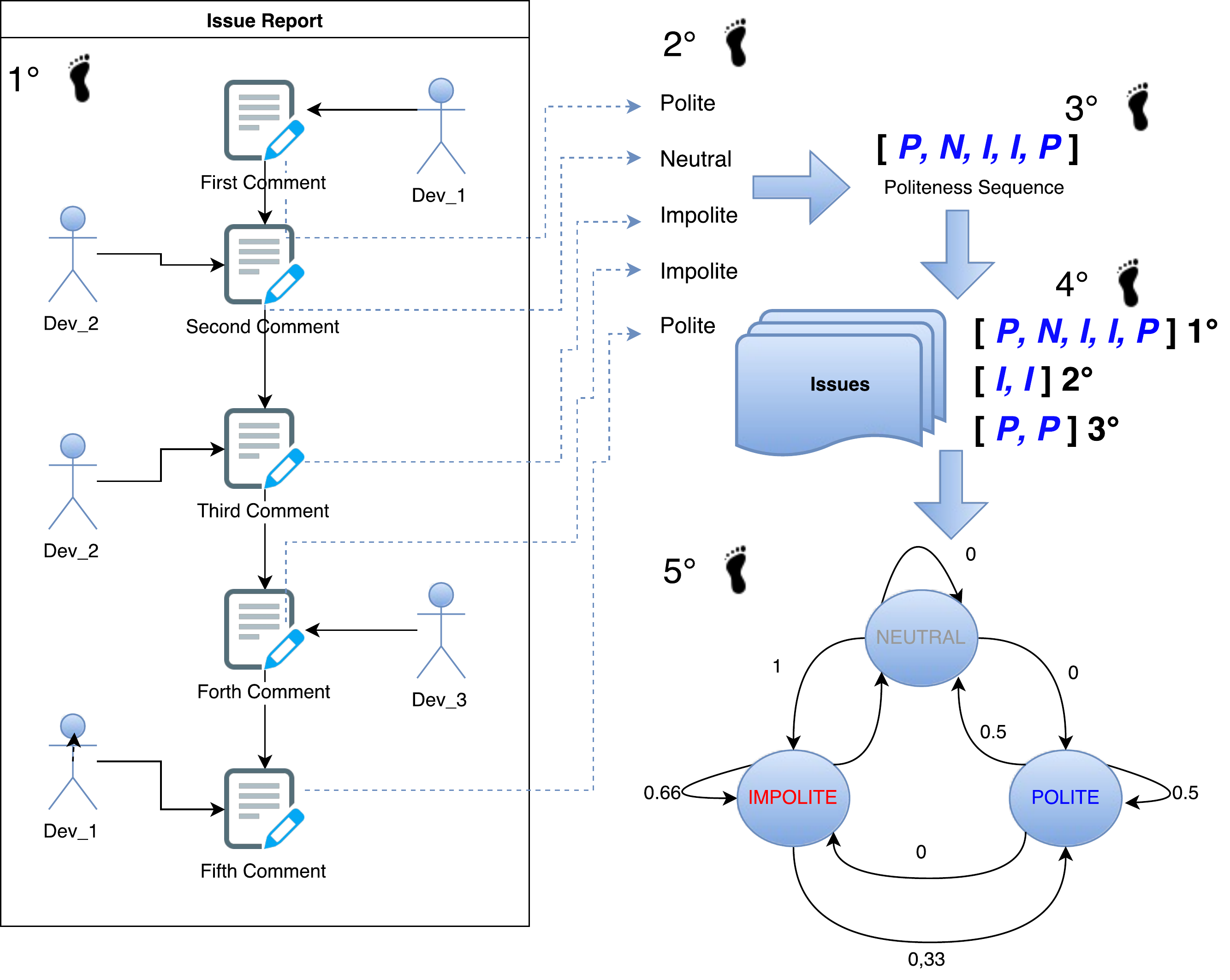}
\caption{Politeness' Markov's chain Schema}
\label{fig:exp_setup_markov}
\end{figure}

We built a MC for each affective metric: sentiment, politeness and
emotion for each project in the corpus. Then, as a second step, we built a comprehensive and general MC for each affective metric considering all the issue reports from the fifteen projects as if they were related to only one project. 
Figure \ref{fig:exp_setup_markov} shows the steps in building
the politeness MC as an example for an issue report in which three developers posted five comments. 

As a first step, we used the politeness tool
\cite{Danescu-Niculescu-Mizil+al:13b} to label each comment as \textit{polite},
\textit{impolite} or \textit{neutral}. 
Next we collected the 
politeness labels of the issue report, considering the set of labels 
as a politeness sequences of \textit{K}-1 pair-wise politeness-transitions ([P,N,I,I,P] in the example), where \textit{K}  is the number of comments in the issue report. 

In this example, the issue report has 4 
transitions: polite-neutral, neutral-impolite, impolite-impolite and
impolite-polite. 
Let us also consider other two transition sequences obtained from other two issues reports, [I,I] and [P,P].
Finally, we counted the frequency of each politeness-transition
obtaining the corresponding MC. 
In our example, if we consider the
\textit{polite} state, we have two transition, P-N and P-P; hence, the 
transition from \textit{polite} to \textit{impolite} state will have a
probability of 0 and the transitions to \textit{polite} and \textit{neutral} state probability 0.5.

The MC for sentiment is built in a similar way to the politeness MC. 
The MC which models emotion transitions is slightly different; however, a comment can be polite, impolite or
neutral when considering politeness, but it might contain more than one emotion. We used the emotion classifier proposed by Ortu et al.
\cite{ortubullies} to analyze each comment and to attribute to it: Anger, Sadness, Joy and/or Love.
For example, if a comment is labeled as containing \textit{anger} and
\textit{sadness} and the next labeled as containing no emotion (\textit{neutral}), then we consider two transitions 
\textit{anger}-\textit{neutral} and \textit{sadness}-\textit{neutral}.

In the appendix section, we present the MCs obtained for every single project of the corpus using the transition matrix explained at the beginning of this section.
\textbf{P} is the transition matrix for Politeness, \textbf{S} for Sentiment and \textbf{E} for Emotions.
The states for \textbf{P, S} and \textbf{E} are the following:

\[P=\begin{bmatrix}
polite&neutral&impolite\\
\end{bmatrix}\]

\[S=\begin{bmatrix}
positive&neutral&negative\\
\end{bmatrix}\]

\[E=\begin{bmatrix}
sadness&anger&joy&love&neutral\\
\end{bmatrix}\]

Given a transition matrix \textbf{P} for Politeness, the element $p_{11}$ indicates the probability of staying in the state \emph{polite}, $p_{12}$ indicates the probability of shifting from the state \emph{polite} to the state \emph{neutral} and $p_{13}$ indicates the probability of shifting from the state \emph{polite} to the state \emph{impolite}.\\

\section{Results}

Existing research has already explored links between
productivity (as measured by issue fixing time) and discrete emotions, sentiment and
politeness~\cite{destefanis2016software,ortubullies}. The dynamic of an issue resolution 
involves complex interactions between different stakeholders such as users,
developer and managers. A model able to describe such interactions could inform in the decision making process.
The underlying assumption is that a model of social interaction can be
used to understand the impact of a certain comment on the whole issue resolution discussion.\\
As presented in Sec. \ref{sec:exp_markov_chains}, we built three MCs for every project in the corpus and
three general MCs for politeness, sentiment and emotions to understand how
developers reacted to impolite/negative comments when they discuss an issue resolution.\\

\textbf{RQ1: Do developers change behaviour in the context of impolite/negative comments?}

\subsection{Politeness}
\begin{figure}[ht!]
\centering
\includegraphics[width=11cm,height=!]{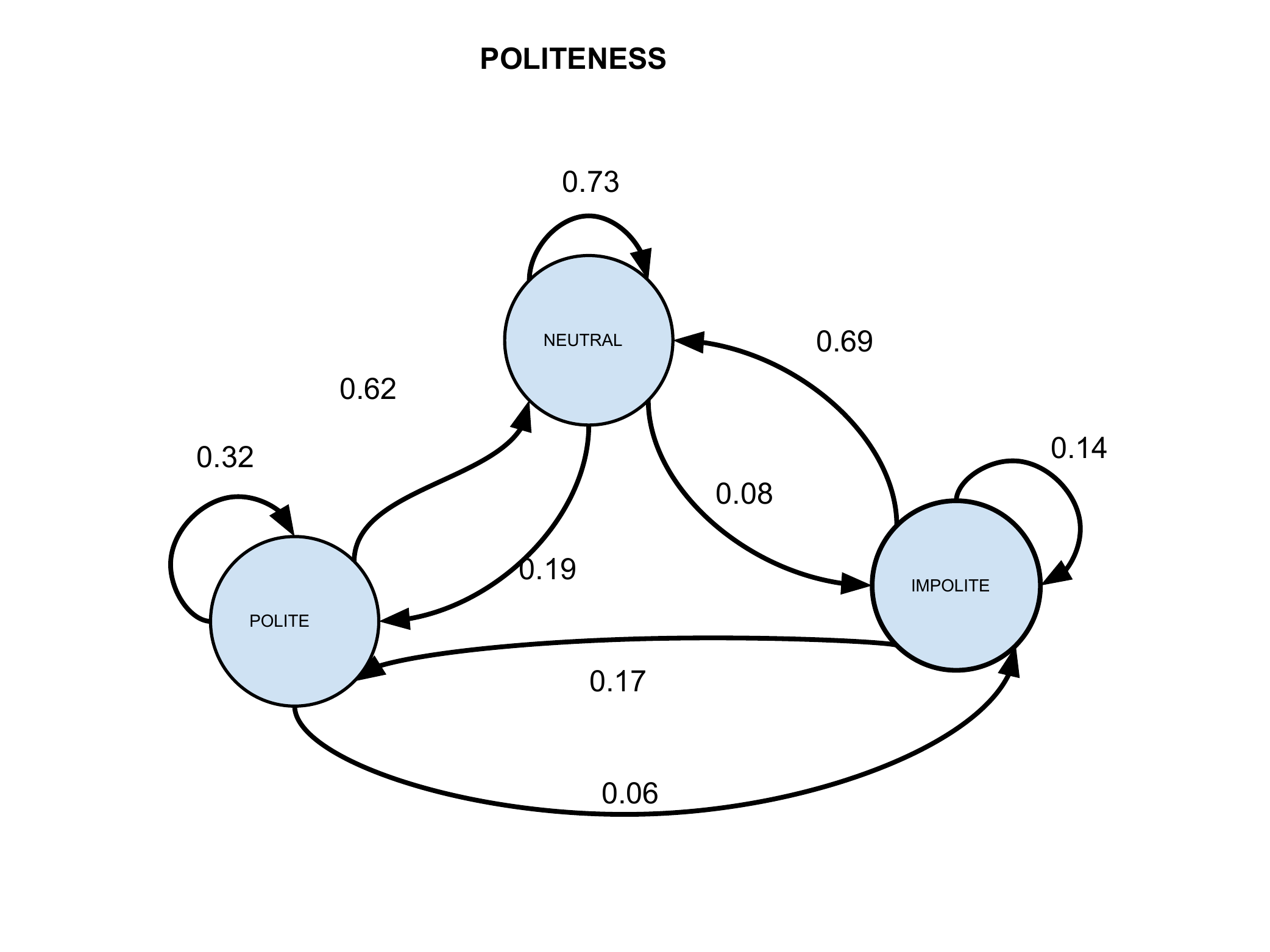}
\caption{Politeness MC for all Projects}
\label{fig:pol_chain}
\end{figure}

\begin{figure*}[htp!]
\centering
\begin{subfigure}{.55\textwidth}
  \centering
   \includegraphics[width=6.6cm]{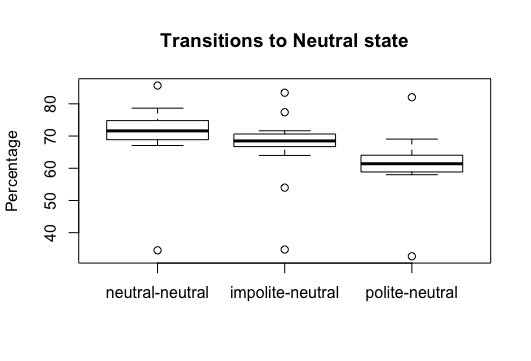}
\end{subfigure}%
\begin{subfigure}{.5\textwidth}
  \centering
  \includegraphics[width=6.6cm]{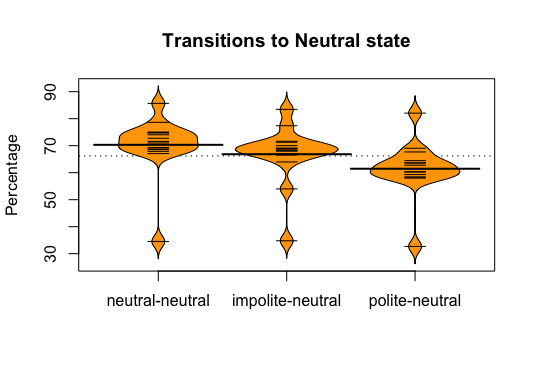}
\end{subfigure}
\caption{Transitions from other states to \textbf{Neutral} state: Boxplot and Beanplot}
\label{fig:trans to Neutral}
\end{figure*}

\begin{figure*}[htp!]
\centering
\begin{subfigure}{.55\textwidth}
  \centering
   \includegraphics[width=6.6cm]{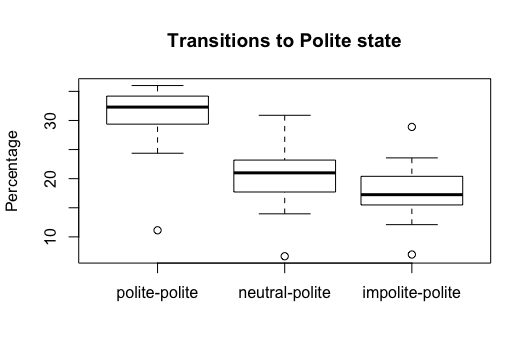}
\end{subfigure}%
\begin{subfigure}{.5\textwidth}
  \centering
  \includegraphics[width=6.6cm]{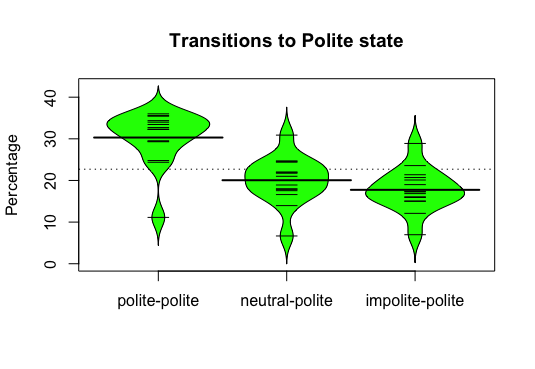}
\end{subfigure}
\caption{Transitions from other states to \textbf{Polite} state: Boxplot and Beanplot}
\label{fig:transToPolite}
\end{figure*}

\begin{figure*}[htp!]
\centering
\begin{subfigure}{.55\textwidth}
  \centering
   \includegraphics[width=6.6cm]{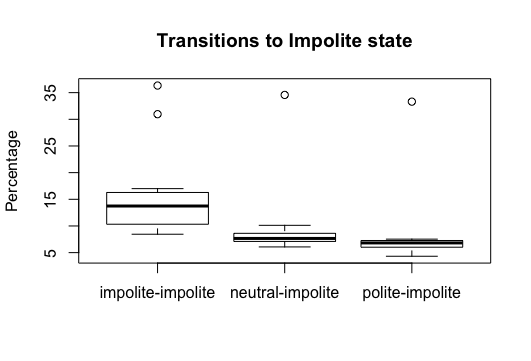}
\end{subfigure}%
\begin{subfigure}{.5\textwidth}
  \centering
  \includegraphics[width=6.6cm]{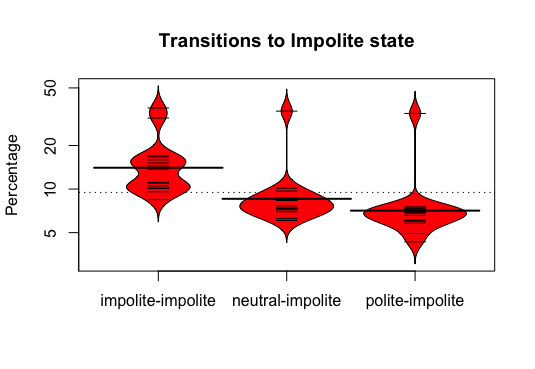}
\end{subfigure}
\caption{Transitions from other states to \textbf{Impolite} state: Boxplot and Beanplot}
\label{fig:transToImpo}
\end{figure*}

Figure \ref{fig:pol_chain} shows the general Politeness' MC describing the probability of changing from a state to another.
Figures \ref{fig:trans to Neutral}, \ref{fig:transToPolite}, \ref{fig:transToImpo}, show the boxplots and beanplots \cite{kampstra2008beanplot} obtained considering all the projects in the corpus. Each single boxplot and beanplot shows the statistics for an array of 15 probabilities (one probability for project).   

\begin{table}[htp!]
\setlength{\tabcolsep}{10pt}
\centering
\caption{Transition Matrix for Politeness MC, with absolute transitions}
\label{tab:pol_chain_abs}
\begin{tabular}{lrrr}
\toprule
	& POLITE & NEUTRAL   & IMPOLITE \\
\midrule
POLITE    & 46049 &	88398 &	8917\\
NEUTRAL   & 23623 &	89969 &	9636\\
IMPOLITE  & 17510 &	71436 &	14349\\
\bottomrule
\end{tabular}
\end{table}

 It is interesting to see that the transition probability values of the general MC (built considering all the issue reports of our corpus together) are quite close to the values of the medians (Figures \ref{fig:trans to Neutral},\ref{fig:transToPolite},\ref{fig:transToImpo}) obtained considering every single project of the corpus.
Table \ref{tab:pol_chain_abs} shows the absolute transitions matrix for Politeness.
The ``neutral'' state is quite stable. If a comment is classified as ``neutral'', communication flow among the developers involved tends to stay neutral, with 
a 73\% probability. There is an 8\% probability of a state-shift from ``neutral'' to ``impolite'' and a 19\% probability of a state-shift from ``neutral'' to ``polite''.
Starting from a ``polite'' state, the probability of shifting to the ``impolite'' state is quite low, 6\%. There is a high probability of moving to the ``neutral'' 
state (62\%). The probability of staying in the same state is 32\%.

Starting from an ``impolite'' state, the probability of moving to a ``polite'' state is 17\%. This is higher than the probability of moving from a ``polite'' 
state to ``impolite''. It is interesting to see that 
the probability of staying in an ``impolite'' state is only 14\% (far lower than the probabilities of staying in both ``neutral'' and ``polite states), 
and that there is a 70\% of probability of a shift from ``impolite'' to ``neutral''. 

Figures \ref{fig:trans to Neutral}, \ref{fig:transToPolite}, \ref{fig:transToImpo}, show the presence of several outliers. 
The projects Wicket and Cassandra are outliers for the transition neutral-neutral, 0.3454 and 0.8567 respectively, while the median for this transition is slightly above 0.7.  
Wicket is an outlier for all the polite transitions:
\[
\mathbf{P_{Wicket}} =\left(
\begin{array}{ccc}
0.34	 & 0.3269 &	0.3331\\
0.3089 &	\textbf{0.3454} &	0.3457\\
0.2889 &	0.3478 &	0.3633
\end{array} \right)
\]
All the probability values look very similar, and transitions from a state to another have quite the same probability.

Cassandra is also outlier for the transitions polite-polite and impolite-neutral. There is only 11.13\% of probability to stay in the Polite state after a Polite comment.
\[
\mathbf{P_{Cassandra}} =\left(
\begin{array}{ccc}
0.1113	&  0.8205  &  	0.0682 \\
0.0667	&  \textbf{0.8567} &  	0.0766 \\
0.0697	&  0.8345   &  0.0958 
\end{array} \right)
\]
For Cassandra, the Neutral state is the one with the highest probabilities.

\subsection{Sentiment}

\begin{figure}[ht!]
\centering
\includegraphics[width=12cm,height=!]{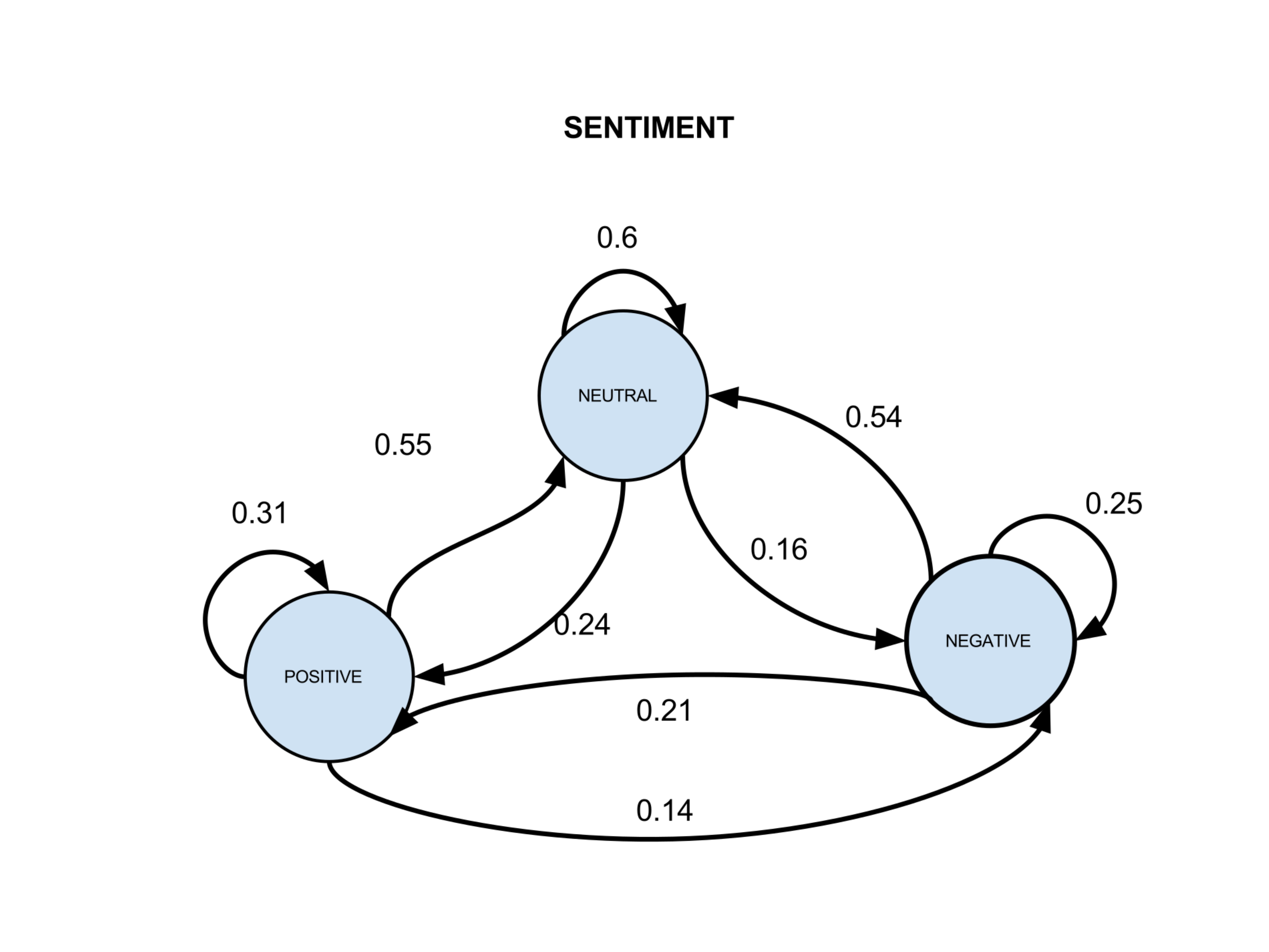}
\caption{Sentiment MC for all Projects}
\label{fig:sent_chain}
\end{figure}

\begin{figure*}[htp!]
\centering
\begin{subfigure}{.49\textwidth}
  \centering
   \includegraphics[width=6.3cm]{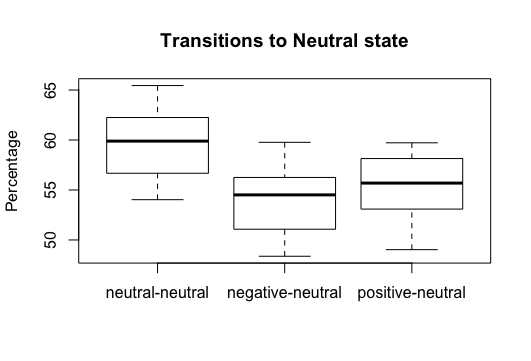}
\end{subfigure}%
\begin{subfigure}{.5\textwidth}
  \centering
  \includegraphics[width=6.4cm]{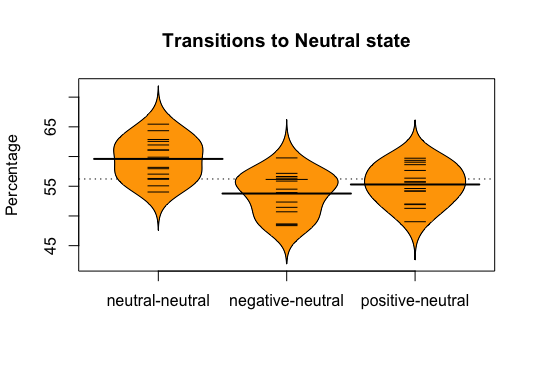}
\end{subfigure}
\caption{Transitions from other states to Neutral state: Boxplot and Beanplot}
\label{fig:transToNeutralS}
\end{figure*}

\begin{figure*}[htp!]
\centering
\begin{subfigure}{0.49\textwidth}
  \centering
   \includegraphics[width=6.3cm]{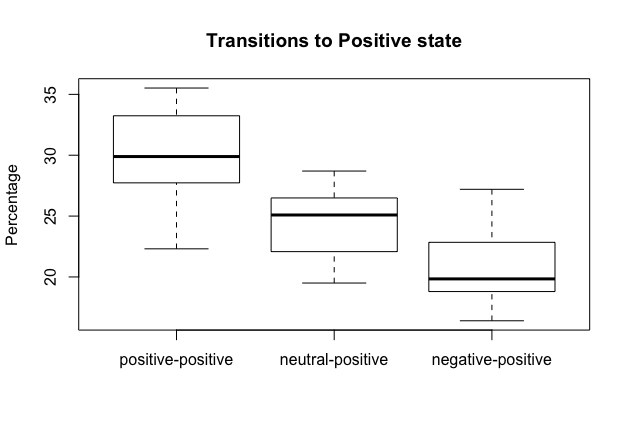}\end{subfigure}
\begin{subfigure}{.5\textwidth}
  \centering
  \includegraphics[width=6.4cm]{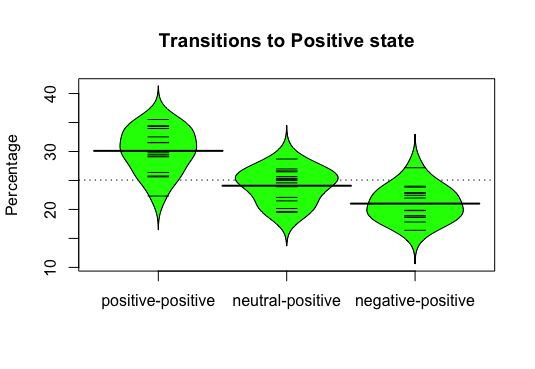}
\end{subfigure}
\caption{Transitions from other states to Positive state: Boxplot and Beanplot}
\label{fig:transToPositive}
\end{figure*}

\begin{figure*}[htp!]
\centering
\begin{subfigure}{.49\textwidth}
  \centering
   \includegraphics[width=6.3cm]{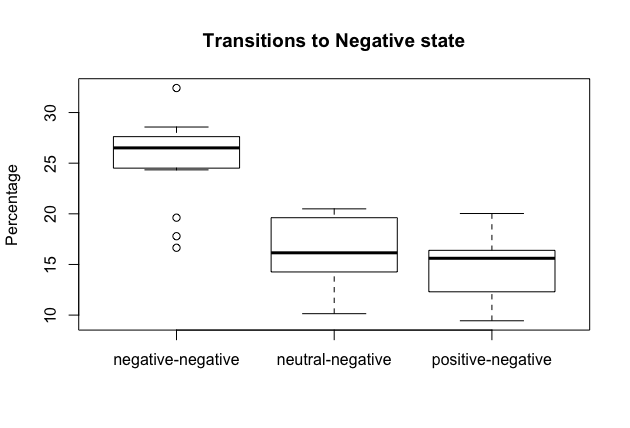}
\end{subfigure}
\begin{subfigure}{.5\textwidth}
  \centering
  \includegraphics[width=6.4cm]{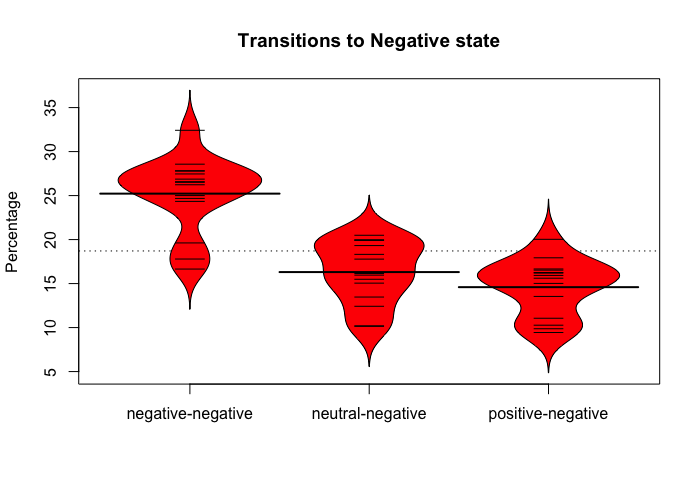}
\end{subfigure}
\caption{Transitions from other states to Negative state: Boxplot and Beanplot}
\label{fig:transToNegative}
\end{figure*}

Figures \ref{fig:transToNeutralS}, \ref{fig:transToPositive}, \ref{fig:transToNegative}, show the boxplots and beanplots obtained considering all the projects in the corpus. Each single boxplot and beanplot shows the statistics for an array of 15 probabilities (one probability for project).  
Figure \ref{fig:sent_chain} shows the general Sentiment MC which describes the
probability of changing from one state to another.

\begin{table}[htp!]
\setlength{\tabcolsep}{10pt}
\centering
\caption{Transition Matrix for Sentiment MC, with absolute transitions}
\label{tab:sent_chain_abs}
\begin{tabular}{lrrr}
\toprule
	& POSITIVE & NEUTRAL   & NEGATIVE \\
\midrule
POSITIVE  & 27050 &	49221  &	12784\\
NEUTRAL   & 66688 &	164318 &	44151\\
NEGATIVE  & 11178 &	28498  &	13331\\
\bottomrule
\end{tabular}
\end{table}

Also in this case, it is interesting to see that the transition probability values of the general MC (built considering all the issue reports of our corpus together) are quite close to the values of the medians (Figures \ref{fig:transToNeutralS}, \ref{fig:transToPositive}, \ref{fig:transToNegative}) obtained considering every single project of the corpus.
Table \ref{tab:sent_chain_abs} shows the absolute transitions matrix for Sentiment.
The ``neutral'' state in this case is also quite stable. If a comment is classified as ``neutral'', communication flow among developers tends to 
stay neutral, with a 60\% probability. There is a 16\% probability of a state-shift from ``neutral'' to ``negative'' and a 24\% probability of a state-shift 
from ``neutral'' to ``positive''.

Starting from a ``positive'' state, the probability of a shift to the ``negative'' state is 14\%. The probability of a move to the ``neutral'' state is 55\%. 
The probability of staying in the same state is 31\%.

From a ``negative'' state, the probability of moving to a ``positive'' state is 21\%. In this case, the value is higher than the probability of moving from a 
``positive'' state to a ``negative'' one. The probability of staying in a ``negative'' state is 25\% (also lower than the probabilities of staying in both 
``neutral'' and ``positive'' states), and that there is a 54\% probability to shift from ``negative'' to ``neutral''.

The outliers occur for the transition negative-negative and are the following systems: Derby, OFBiz, Infrastructure and Camel. 

For Derby, the probability of staying in the negative state (after a negative) is higher than 30\%, while for OFBiz, Infrastructure and Camel is lower than 20\%, as shown in the following transition matrices.

\[ 
\mathbf{S_{Derby}} = \left( \begin{array}{lcr}
0.3252 &	0.5128	& 0.1619\\
0.2392	& 0.5617 &	0.1991\\
0.1892 &	0.4866 &	\textbf{0.3242}
\end{array} \right)
\mathbf{S_{OFBiz}} = \left( \begin{array}{lcr}
0.3396 &	0.5577 &	0.1027\\
0.2699 &	0.6287 &	0.1014\\
0.272 &	0.5615 &	\textbf{0.1665}
\end{array} \right)
\]
\[
\mathbf{S_{Camel}} = \left( \begin{array}{lcr}
0.3151 &	0.5905 &	0.0944\\
0.287 &	0.611 &	0.102\\
0.2244 &	0.5977 &	\textbf{0.1779}
\end{array} \right)
\]

\textbf{RQ2: What is the probability of shifting from
comments holding positive emotions to comments holding negative emotion?}\\

The first research question showed how developers tended to respond more
positively than negatively when considering politeness and sentiment. It is interesting to analyze if the same behaviours occur for emotions.\\
We built the MCs for emotions as presented in
Sec. \ref{sec:exp_markov_chains} to analyze the probabilities of
shifting from an emotion to another when developers communicate.\\

\begin{table}[htp!]
\setlength{\tabcolsep}{10pt}
\centering
\caption{Transition Matrix for Emotion MC}
\label{tab:emo_chain}
\begin{tabular}{lrrrrr}
\toprule
        & SADNESS & ANGER   & JOY     & LOVE    & NEUTRAL \\
\midrule
SADNESS & 26.11\% & 4.49\%  & 7.88\%  & 6.45\%  & 55.08\% \\
ANGER   & 13.79\% & 40.11\% & 5.61\%  & 4.10\%  & 36.39\% \\
JOY     & 17.46\% & 4.43\%  & 11.89\% & 12.22\% & 54.00\% \\
LOVE    & 15.84\% & 3.84\%  & 8.29\%  & 15.59\% & 56.44\% \\
NEUTRAL & 16.42\% & 4.29\%  & 7.64\%  & 7.80\%  & 63.85\% \\
\bottomrule
\end{tabular}
\end{table}

\begin{table}[htp!]
\setlength{\tabcolsep}{10pt}
\centering
\caption{Transition Matrix for Emotion MC, with absolute transitions}
\label{tab:emo_chain_abs}
\begin{tabular}{lrrrrr}
\toprule
	& SADNESS & ANGER   & JOY     & LOVE    & NEUTRAL \\
\midrule
SADNESS & 19148	  & 3293    & 5779    & 4731    & 40396\\
ANGER   & 16390   & 47674   & 6664    &	4873	& 43257\\
JOY     & 6385    & 1620    & 4347    &	4470    & 19748\\
LOVE    & 3490    & 845     & 1826    &	3436    & 12433\\
NEUTRAL & 41142   & 10749   & 19147   &	19547   & 160018\\
\bottomrule
\end{tabular}
\end{table}

\begin{figure}[htp!]
\centering
\includegraphics[width=12cm,height=!]{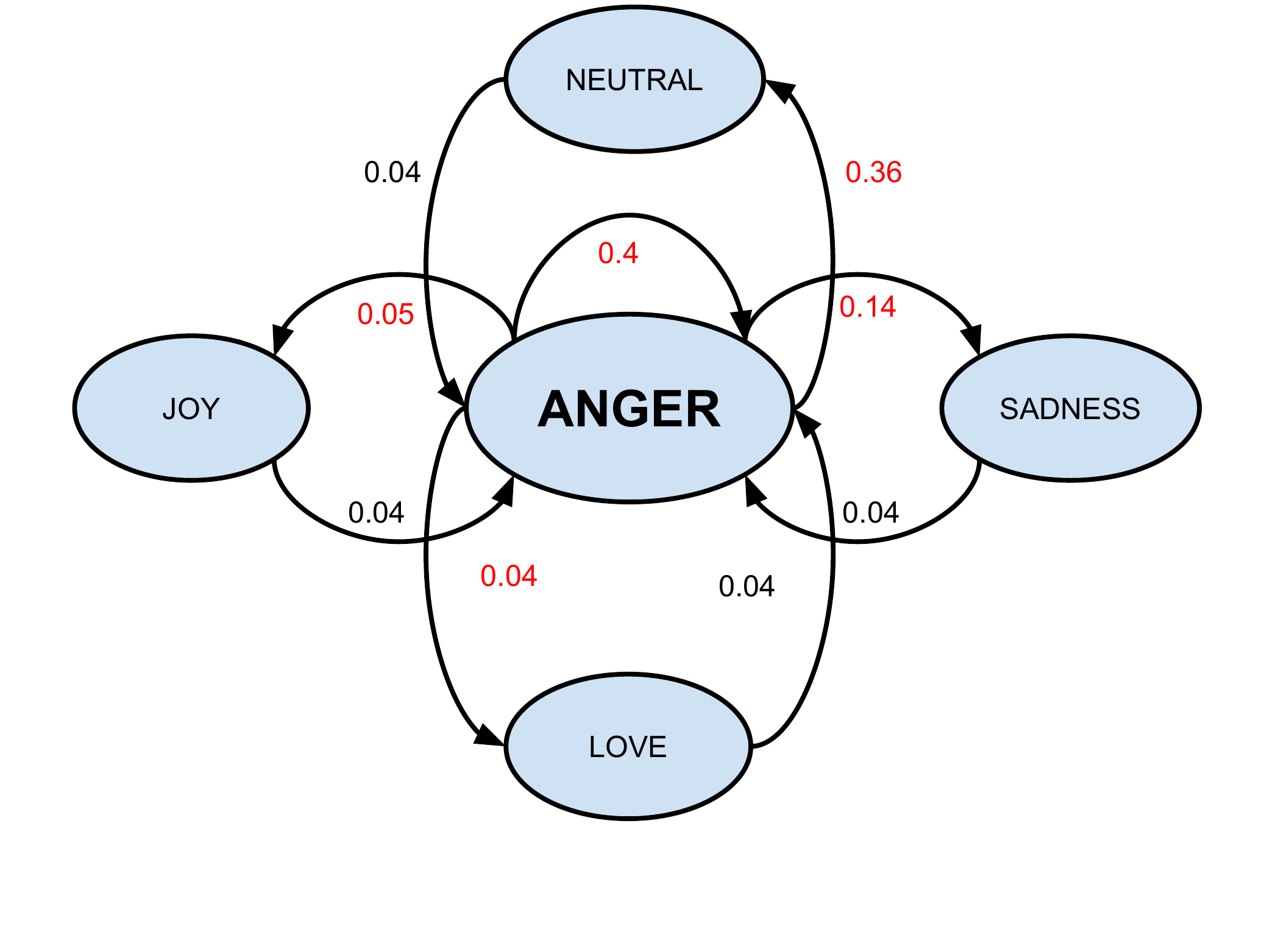}
\caption{Anger Markov chain. For simplicity only edges from/to anger are
diplayed}
\label{fig:emo_chain}
\end{figure}

Negative emotions such as \textit{sadness} and
\textit{anger} tend to be followed by negative emotions more than positive
emotion are followed by positive emotions.
Table \ref{tab:emo_chain} shows the general emotion transition matrix (a single emotion transition matrix E, for each project in the corpus, is presented in the Appendix section). 
Table \ref{tab:emo_chain_abs} shows the absolute transitions matrix.
As for previous
MCs, the numbers represent the probability of a comment containing emotion X being
followed by a comment containing emotion Y (e.g., a comment expressing
\textit{sadness} has a probability of 26.11\% of being followed by another
\textit{sadness} comment).

As confirmed by other studies \cite{Murgia:2014}, most of the
comments expressing emotion are likely to be followed by \textit{neutral}
comments, with the exception of \textit{anger}.
Figure \ref{fig:emo_chain} is a graphical representation of the portion of Table
\ref{tab:emo_chain} for the \textit{anger} emotion showing it has probability of 40\% of being followed by an \textit{anger} comment against probability of 36\% to be followed by a \textit{neutral} comment.

\section{Discussion}
\label{sec:discussion} 
In this paper, we presented MC models describing how developers interacted with each other, analyzing comments posted on an issue tracking system. We selected the 15 most commented projects from the Jira issue report dataset of Ortu et al. \cite{ortu2015jiradataset,ortu2016jiraMSR}, which contained 700,000 issues from 1000 open source projects and two million issue comments. Building software (either for a company or volunteering for an open source project) is, nowadays, a collaborative activity. Activities in which there are groups of people working together to reach a given goal, need structure and coordination. Tools such as issue tracking systems provide enormous help in managing activities and people and such tools are becoming valuable sources of information for both managers and researchers. Furthermore, the study of ``behavioural software engineering''  \cite{lenberg2015behavioral} is gaining increasing importance as a key factor for improving (in every sense and direction) the software development process. Conflicts affect developer productivity and managers are certainly interested in knowing how to prevent, avoid, or, in the worst case, manage conflicts which might occur. We knew from previous studies that information mined from software repositories, like issue repositories, contained emotions \cite{Murgia:2014} and could offer a way to investigate productivity during the software development process \cite{destefanis2016software,ortubullies,ortu2015would}. 

The premise for this study was that productivity had been shown to be higher when developers were motivated, polite and work in a good ``frame of mind'' \cite{destefanis2016software,graziotin2014happy,ortubullies}. Therefore, we wanted to build a baseline model of ``what is going on'' in a generic development environment. The 15 projects in our study were selected looking only at the number of comments posted by developers on the issue tracking system, and without any other specific reason related to affectiveness management. We had no pre-conceived notions of what kind of results we would obtain and wanted to understand if a “self adjustment” toward higher productivity was in the nature of developer interaction. More generally, we wanted to understand the starting point for affectiveness in a software development environment, considering that politeness and good manners can be related to higher productivity.

Considering politeness, we found that the probability of moving to a \textbf{polite} state from an \textbf{impolite} state was higher than \textit{vice versa}  (17\% against 6\%). Also, for sentiment, we found that the probability of moving to a \textbf{positive} state from a \textbf{negative} state was higher than the probability of moving from a \textbf{positive} state to a \textbf{negative} (21\% against 14\%). From a managerial point of view, both \textbf{polite} and \textbf{positive} states should be preferred (because they are related to higher productivity) and actions should be taken to encourage and help all people involved in the development process to stay in those two states. An example of those actions could be related to (but not limited to) an optimized communication environment aiming at conflict minimization, a balanced workload for each developer, updated workstations and an appealing working environment.

We also found several outliers for both politeness and sentiment. As highlighted in Section 5.1, Wicket and Cassandra were outliers for the transition: neutral-neutral; Wicket was also an outlier for all polite transitions and Cassandra for the transitions: polite-polite and impolite-neutral. Derby, OFBiz, Infrastructure and Camel were outliers for the transition: negative-negative (see Section 5.2). We manually investigated the issue tracking system of the outlier projects and several factors could be held responsible. One factor could be the different distributions of JIRA maintenance types and issue priorities. Destefanis et al. \cite{destefanis2016software} showed that different issue priorities and maintenance types were related to different levels of politeness. Bug issue typologies were those with lower politeness. When something is broken and needs to be fixed, the situation is less conducive to politeness and could consequently generate impolite reactions. On the other hand, New Feature issue typologies were those with higher politeness; Trivial issue typologies were those characterised by lower politeness (these might be related to minor programming mistakes and/or poor knowledge of programming practices). A further   factor could be related to the number of developers involved in the projects and the workload distributed among those developers. Ortu et al. \cite{ortujira} highlighted the presence of Pareto’s law in projects developed using Jira (20\% of developers undertaking 80\% of the issue resolution) and showed that there were only a few communities taking care of the majority of issues. OFbiz, Infrastructure, Camel and Wicket also have a similar (lower) number of comments and this could have affected our analysis.  

Considering emotions, we found that the probability of moving from the anger state to positive states like joy and love, was quite low (5\% and 4\% respectively), while there was a very high probability of staying in the anger state (40\%) and moving to the neutral state (36\%). The probability of moving from anger to sadness was 14\% and this result shows that, in our corpus, it is more difficult to drift away from negative emotions. The  highly cited paper of Baumeister et. al \cite{baumeister2001bad}, does not find a counter-example to the results we obtained considering emotions.
Several studies have showed how emotional states can be transferred to others; the spread of emotions was studied using data from a large social network collected over a 20 year period, suggesting that longer-lasting moods such as happiness and depression can be transferred over networks \cite{fowler2008dynamic}.

Fan et al. \cite{fan2016easier} analyzed millions of tweets in Weibo, a Twitter-like service in China, finding that anger was more contagious than joy. Kramer et al. \cite{kramer2014experimental} performed an experiment with people on Facebook (the way in which developers interact on Jira has similarities with Facebook) and tested whether emotional contagion occurred outside of in-person interaction between individuals by reducing the amount of emotional content in a News Feed. When positive expressions were reduced, people produced fewer positive posts and more negative posts; when negative expressions were reduced, the opposite trend occurred.  The  results also suggested that in-person interaction and non-verbal cues were not strictly necessary for emotional contagion. Ferrara et al. \cite{ferrara2015quantifying} used Twitter as a case study and found that negative messages spread faster than positive ones, but positive ones reached larger audiences, suggesting that people are more inclined to share  positive contents, the so-called positive bias.

Predicting a shift toward impolite or negative states can help managers in taking actions aimed at keeping the general mood high and relaxed, lowering and preventing conflicts and obtaining higher productivity as a result. The results presented in this study can be also used as a baseline for comparing the ``as-is'' state of other projects/companies. Models such as MCs presented in this study could also be helpful when defining teams of developers. Knowing the profile (from a politeness point of view) of developers might provide hints for creating more team balance.

\section{Threats to validity}
Several threats to validity need to be considered. Threats to external validity are related to generalisation of
our conclusions. With regard to the system studied in this work, we
considered only open-source systems and this could affect
the generality of the study; our results are not meant to be
representative of all environments or programming languages.
Commercial software is typically developed using different platforms and technologies, 
with strict deadlines and cost limitations and by developers with different experience. 
Politeness, sentiment and emotions measures are approximations given the
challenges of natural language and subtle phenomena like sarcasm. To deal with
these threats, we used SentiStrength form measuring sentiment, Danescu et
al.'s politeness tool \cite{Danescu-Niculescu-Mizil+al:13b} and Ortu et al.
\cite{ortubullies} for measuring politeness. While Danescu et al.'s politeness tool \cite{Danescu-Niculescu-Mizil+al:13b} has been trained using Stack Overflow, hence is reliable in the software engineering domain, SentiStreght has not been trained on software engineering data and its application in the software engineering context can be problematic\cite{jongeling2015choosing,novielli2015challenges}.
 
Threats to internal validity concern confounding factors that could influence the obtained results.
Since the comments used in this study were collected over an extended period from developers 
unaware of being subject to analysis, 
we are confident that the emotions we mined are genuine. 
Confounds could have affected validity of the results for our research questions, since the number of developers involved in discussing issues might differ, as well as severity and complexity of an issue under analysis. 
This study is focused on text written by agile developers for developers. To
correctly depict the affectiveness embedded in such comments, it is necessary to
understand the developers' dictionary and slang. 
This assumption is supported by Murgia et al. \cite{Murgia:2014} for
measuring emotions.

\section{Conclusions and future work}
\label{Conclusion}
This paper presented an analysis of more than 500K comments 
from open-source issue tracking system repositories. We empirically determined
how developers interacted with each other under certain psychological conditions generated 
by politeness, sentiment and emotions of a comment posted on a issue tracking
system.
Results showed that when in the presence of impolite or negative comments, there is higher 
probability for the next comment to be neutral or polite (neutral or positive in case of sentiment) than impolite or negative.
This fact demonstrates that developers, in the dataset considered for this study, tended to resolve 
conflicts instead of increasing negativity within the communication flow. This
is not true when we consider emotions; negative emotions are more likely to be
followed by negative emotions than positive.\\ 
Markov models provide a mathematical description of developer behavioural aspects and the result could help managers take control the development phases of a system (expecially in a distributed environment), since social aspects can seriously affect a developer's productivity. 
As future works we plan to investigate possible links existing between software metrics and emotions, to better understand the impact of affectiveness on software quality.

\section{Acknowledgement}
The research presented in this paper was partly funded by the  Engineering and Physical Sciences Research Council (EPSRC) of the UK under grant ref: EP/M024083/1.

\section{Appendix}
In this section we present all the matrices for all the fifteen systems in our corpus.
\textbf{P} is the transition matrix for Politeness, \textbf{S} for Sentiment and \textbf{E} for Emotions.
The states for \textbf{P, S} and \textbf{E} are the following:

\[P=\begin{bmatrix}
polite&neutral&impolite\\
\end{bmatrix}\]

\[S=\begin{bmatrix}
positive&neutral&negative\\
\end{bmatrix}\]

\[E=\begin{bmatrix}
sadness&anger&joy&love&neutral\\
\end{bmatrix}\]

The following matrices are the transition matrices for Politeness, Sentiment and Emotion for HBase.

\[ \mathbf{P_{HBase}} =\left(
      \begin{array}{lcr}
0.3594 & 0.5832 & 0.0574\\
0.218 & 0.7077 & 0.0743\\
0.1926 & 0.7036 & 0.1038
      \end{array} \right)
\mathbf{S_{HBase}} = \left( \begin{array}{lcr}
0.2576	& 0.5863 &	0.1561\\
0.2016	& 0.6435 &	0.1549\\
0.1783	&0.5717	& 0.25
\end{array} \right)
\]
\[
\mathbf{E_{HBase}} = \left( \begin{array}{ccccc}
0.203 &	0.2063 &	0.0686 &	0.0486 &	0.4736\\
0.127 &	0.4388 & 0.0508 &	0.0326 &	0.3508\\
0.1414 &	0.179 &	0.1578 &	0.0562 &	0.4656\\
0.1396 &	0.1787 &	0.0856 &	0.1243 &	0.4719\\
0.1349 &	0.1999 &	0.0691 &	0.0507 &	0.5454
\end{array} \right)
\]

The following matrices are the transition matrices for Politeness, Sentiment and Emotion for Hadoop Common.

\[ \mathbf{P_{HCommon}} =\left(
      \begin{array}{lcr}
0.293	&  0.6363  &  	0.0707 \\
 0.189	&  0.7275	 &  	0.0834 \\
 0.1725	&  0.7163   &  0.1112
      \end{array} \right)
\mathbf{S_{HCommon}} = \left( \begin{array}{lcr}
0.2953	& 0.5421 &	0.1626\\
0.2519 &	0.5704 &	0.1778\\
0.2197 &	0.5145 &	0.2658
\end{array} \right)
\]
\[
\mathbf{E_{HCommon}} = \left( \begin{array}{ccccc}
0.2716 &	0.0127 &	0.0875 &	0.0695 &	0.5587\\
0.2568 &	0.0309 & 0.0927 &	0.0888 &	0.5309\\
0.1593 &	0.0095 &	0.0989 &	0.1542 &	0.5781\\
0.1608 &	0.0081 &	0.0801 &	0.1283 &	0.6227\\
0.1638 &	0.008 &	0.0858 &	0.1042 &	0.6382
\end{array} \right)
\]

The following matrices are the transition matrices for Politeness, Sentiment and Emotion for Derby.

\[ \mathbf{P_{Derby}} =\left(
 \begin{array}{ccc}
0.3544	&  0.6024  &  	0.0432 \\
 0.2468	&  0.6925	 &  	0.0607 \\
 0.2138	&  0.6848   &  0.1014 
\end{array} \right)
\mathbf{S_{Derby}} = \left( \begin{array}{lcr}
0.3252 &	0.5128	& 0.1619\\
0.2392	& 0.5617 &	0.1991\\
0.1892 &	0.4866 &	0.3242
\end{array} \right)
\]
\[
\mathbf{E_{Derby}} = \left( \begin{array}{ccccc}
0.288 & 0.0097 & 0.0723 & 0.0931 & 0.5369\\
0.2396 & 0.0685 & 0.0954 & 0.0954 & 0.5012\\
0.2206 & 0.0126 & 0.0974 & 0.1705 & 0.4989\\
0.218 & 0.0064 & 0.0807 & 0.1554 & 0.5395\\
0.2152 & 0.0067 & 0.07 & 0.0824 & 0.6257
\end{array} \right)
\]

The following matrices are the transition matrices for Politeness, Sentiment and Emotion for Lucene Core.

\[ \mathbf{P_{LCore}} =\left(
\begin{array}{ccc}
        0.3558	&  0.5839  &  	0.0603 \\
 0.2439	&  0.6706 &  	0.0854 \\
0.2011	&  0.6398   &  0.1591 
      \end{array} \right)
\mathbf{S_{LCore}} = \left( \begin{array}{lcr}
0.2908 &	0.5636 &	0.1456\\
0.2148 &	0.6255 &	0.1596\\
0.1979 &	0.5587 &	0.2434
\end{array} \right)
\]
\[
\mathbf{E_{LCore}} = \left( \begin{array}{ccccc}
0.3157 & 0.0378 & 0.1009 & 0.0479 & 0.4977\\
0.2578 & 0.1858 & 0.1202 & 0.0389 & 0.3972\\
0.2131 & 0.0331 & 0.1566 & 0.0765 & 0.5206\\
0.1969 & 0.0263 & 0.1057 & 0.1139 & 0.5572\\
0.1975 & 0.0287 & 0.1065 & 0.0541 & 0.6132
\end{array} \right)
\]

The following matrices are the transition matrices for Politeness, Sentiment and Emotion for Hadoop HDFS.

\[ \mathbf{P_{H-HDFS}} =\left(
\begin{array}{ccc}
0.2944	&  0.6446  &  	0.061 \\
0.18	&  0.7504 &  	0.0697 \\
 0.1737	&  0.6816   &  0.1447
\end{array} \right)
\mathbf{S_{H-HDFS}} = \left( \begin{array}{lcr}
0.3431 &	0.4903	& 0.1666\\
0.2665 &	0.5403	& 0.1932\\
0.229 &	0.4853 &	0.2857
\end{array} \right)
\]
\[
\mathbf{E_{H-HDFS}} = \left( \begin{array}{ccccc}
0.2611 & 0.011 & 0.1126 & 0.063 & 0.5523\\
0.2654 & 0.0192 & 0.1 & 0.0654 & 0.55\\
0.1579 & 0.0073 & 0.1206 & 0.1699 & 0.5443\\
0.149 & 0.0033 & 0.1227 & 0.1301 & 0.5949\\
0.1628 & 0.0061 & 0.1171 & 0.0793 & 0.6347
\end{array} \right)
\]

The following matrices are the transition matrices for Politeness, Sentiment and Emotion for Cassandra.

\[ \mathbf{P_{Cassandra}} =\left(
\begin{array}{ccc}
0.1113	&  0.8205  &  	0.0682 \\
0.0667	&  0.8567 &  	0.0766 \\
0.0697	&  0.8345   &  0.0958 
\end{array} \right)
\mathbf{S_{Cassandra}} = \left( \begin{array}{lcr}
0.2231 &	0.5766 &	0.2003\\
0.1962 &	0.5989 &	0.2049\\
0.164 &	0.5615 &	0.2746
\end{array} \right)
\]
\[
\mathbf{E_{Cassandra}} = \left( \begin{array}{ccccc}
0.2838 & 0.0126 & 0.0583 & 0.0245 & 0.6208\\
0.2308 & 0.0398 & 0.0584 & 0.0292 & 0.6419\\
0.2073 & 0.0072 & 0.0945 & 0.0287 & 0.6623\\
0.223 & 0.0068 & 0.0689 & 0.0568 & 0.6446\\
0.1923 & 0.0093 & 0.0605 & 0.0274 & 0.7104
\end{array} \right)
\]

The following matrices are the transition matrices for Politeness, Sentiment and Emotion for Solr.

\[ \mathbf{P_{Solr}} =\left(
\begin{array}{ccc}
0.3226 &	0.6036 &	0.0738\\
0.2197 &	0.6791 &	0.1012\\
0.1684 &	0.6651 &	0.1665
\end{array} \right)
\mathbf{S_{Solr}} = \left( \begin{array}{lcr}
0.2562 &	0.5936 &	0.1502\\
0.195 &	0.6545 & 	0.1505\\
0.1868 &	0.5664 &	0.2468
\end{array} \right)
\]
\[
\mathbf{E_{Solr}} = \left( \begin{array}{ccccc}
0.296 & 0.0131 & 0.0792 & 0.0453 & 0.5665\\
0.2884 & 0.0344 & 0.0794 & 0.0476 & 0.5503\\
0.216 & 0.0137 & 0.1284 & 0.0612 & 0.5806\\
0.1622 & 0.0056 & 0.0783 & 0.085 & 0.6689\\
0.1792 & 0.0089 & 0.0682 & 0.0533 & 0.6904
\end{array} \right)
\]

The following matrices are the transition matrices for Politeness, Sentiment and Emotion for Hive.

\[ \mathbf{P_{Hive}} =\left(
\begin{array}{ccc}
0.3344 &	0.593 &	0.0726\\
0.1755 &	0.7404 &	0.0841\\
0.1591 &	0.6896 &	0.1513
\end{array} \right)
\mathbf{S_{Hive}} = \left( \begin{array}{lcr}
0.3445 & 0.5201 &	0.1354\\
0.2458 &	0.6195 &	0.1347\\
0.1862 &	0.5451 &	0.2687
\end{array} \right)
\]
\[
\mathbf{E_{Hive}} = \left( \begin{array}{ccccc}
0.2165 & 0.0793 & 0.0778 & 0.0724 & 0.5541\\
0.0856 & 0.4896 & 0.0446 & 0.0572 & 0.323\\
0.1294 & 0.06 & 0.0902 & 0.2301 & 0.4904\\
0.1117 & 0.1141 & 0.0759 & 0.1558 & 0.5425\\
0.1169 & 0.0842 & 0.0801 & 0.1091 & 0.6096
\end{array} \right)
\]

The following matrices are the transition matrices for Politeness, Sentiment and Emotion for Hadoop Map/Reduce.

\[ \mathbf{P_{H-M/R}} =\left(
\begin{array}{ccc}
0.327 &	0.615 &	0.058\\
0.212 &	0.716 &	0.072\\
0.208 &	0.691 &	0.101 
\end{array} \right)
\mathbf{S_{H-M/R}} = \left( \begin{array}{lcr}
0.3156 &	0.5191 &	0.1653\\
0.2537 &	0.5631 &	0.1832\\
0.2385 &	0.4837 &	0.2778
\end{array} \right)
\]
\[
\mathbf{E_{H-M/R}} = \left( \begin{array}{ccccc}
0.2632 & 0.0134 & 0.0852 & 0.0783 & 0.5599\\
0.2477 & 0.052 & 0.0734 & 0.0917 & 0.5352\\
0.1568 & 0.0095 & 0.1093 & 0.1901 & 0.5344\\
0.1604 & 0.0092 & 0.0776 & 0.1682 & 0.5846\\
0.161 & 0.0093 & 0.0929 & 0.1052 & 0.6317
\end{array} \right)
\]

The following matrices are the transition matrices for Politeness, Sentiment and Emotion for Harmony.

\[ \mathbf{P_{Harmony}} =\left(
\begin{array}{ccc}
0.2437 &	0.6905 &	0.0658\\
0.1396 &	0.7865 &	0.0739\\
0.121 &	0.7739 &	0.1051
\end{array} \right)
\mathbf{S_{Harmony}} = \left( \begin{array}{lcr}
0.3552	& 0.5462 &	0.0987\\
0.2567 &	0.5818 &	0.1615\\
0.1983 &	0.5234 &	0.2783
\end{array} \right)
\]
\[
\mathbf{E_{Harmony}} = \left( \begin{array}{ccccc}
0.1961 & 0.0145 & 0.0477 & 0.1176 & 0.6241\\
0.1289 & 0.2607 & 0.0344 & 0.0831 & 0.4928\\
0.1265 & 0.0123 & 0.0835 & 0.1265 & 0.6511\\
0.0977 & 0.0051 & 0.0615 & 0.297 & 0.5386\\
0.1435 & 0.0117 & 0.0548 & 0.1357 & 0.6544
\end{array} \right)
\]

The following matrices are the transition matrices for Politeness, Sentiment and Emotion for OFBiz.
\[ \mathbf{P_{OFBiz}} =\left(
\begin{array}{ccc}
0.3437 &	0.5833 &	0.0729\\
0.1787 &	0.7478 &	0.0734\\
0.1506 &	0.5399 &	0.3096 
\end{array} \right)
\mathbf{S_{OFBiz}} = \left( \begin{array}{lcr}
0.3396 &	0.5577 &	0.1027\\
0.2699 &	0.6287 &	0.1014\\
0.272 &	0.5615 &	0.1665
\end{array} \right)
\]
\[
\mathbf{E_{OFBiz}} = \left( \begin{array}{ccccc}
0.275 & 0.0091 & 0.0793 & 0.1085 & 0.5282\\
0.262 & 0.0374 & 0.0963 & 0.1283 & 0.4759\\
0.2286 & 0.0062 & 0.0885 & 0.1592 & 0.5175\\
0.1754 & 0.0086 & 0.0809 & 0.2263 & 0.5088\\
0.1916 & 0.0094 & 0.0668 & 0.1533 & 0.5789
\end{array} \right)
\]

The following matrices are the transition matrices for Politeness, Sentiment and Emotion for Infrastracture.
\[ \mathbf{P_{Inf}} =\left(
\begin{array}{ccc}
0.2479 &	0.6768 &	0.0753\\
0.1661 &	0.747 &	0.087\\
0.1499 &	0.7127 &	0.1374 
\end{array} \right)
\mathbf{S_{Inf}} = \left( \begin{array}{lcr}
0.2922 &	0.5972 &	0.1106\\
0.2649 &	0.6109 &	0.1242\\
0.2401 &	0.5636 &	0.1962
\end{array} \right)
\]
\[
\mathbf{E_{Inf}} = \left( \begin{array}{ccccc}
0.2388 & 0.0069 & 0.0659 & 0.0862 & 0.6022\\
0.2407 & 0.0093 & 0.037 & 0.1111 & 0.6019\\
0.1776 & 0.0088 & 0.1151 & 0.1346 & 0.5639\\
0.1499 & 0.0035 & 0.0826 & 0.1194 & 0.6447\\
0.181 & 0.0061 & 0.0645 & 0.1166 & 0.6319
\end{array} \right)
\]
The following matrices are the transition matrices for Politeness, Sentiment and Emotion for Camel.
\[ \mathbf{P_{Camel}} =\left(
\begin{array}{ccc}
0.2949 &	0.6359 &	0.0692\\
0.2174 &	0.6857 &	0.0969\\
0.1607 &	0.6693 &	0.17 
\end{array} \right)
\mathbf{S_{Camel}} = \left( \begin{array}{lcr}
0.3151 &	0.5905 &	0.0944\\
0.287 &	0.611 &	0.102\\
0.2244 &	0.5977 &	0.1779
\end{array} \right)
\]
\[
\mathbf{E_{Camel}} = \left( \begin{array}{ccccc}
0.1651 & 0.078 & 0.0746 & 0.0698 & 0.6125\\
0.0902 & 0.5499 & 0.0438 & 0.0396 & 0.2765\\
0.1213 & 0.0936 & 0.0893 & 0.0841 & 0.6118\\
0.1297 & 0.0692 & 0.0701 & 0.1076 & 0.6234\\
0.1175 & 0.0586 & 0.0634 & 0.0707 & 0.6898
\end{array} \right)
\]
The following matrices are the transition matrices for Politeness, Sentiment and Emotion for ZooKeeper.
\[ \mathbf{P_{ZooK}} =\left(
\begin{array}{ccc}
0.323 &	0.6278 &	0.0493\\
0.2459 &	0.6916 &	0.0625\\
0.2357 &	0.6798 &	0.0845
\end{array} \right)
\mathbf{S_{ZooK}} = \left( \begin{array}{lcr}
0.2989 &	0.5417 &	0.1593\\
0.2499 &	0.5507 &	0.1994\\
0.2279 &	0.507 &	0.2651
\end{array} \right)
\]
\[
\mathbf{E_{ZooK}} = \left( \begin{array}{ccccc}
0.158 & 0.0008 & 0.0145 & 0.0015 & 0.8252\\
0.1333 & 0.0667 & 0 & 0 & 0.8\\
0.1588 & 0.0059 & 0.0471 & 0.0059 & 0.7824\\
0.0714 & 0 & 0 & 0.0714 & 0.8571\\
0.0987 & 0.0008 & 0.0134 & 0.0011 & 0.886
\end{array} \right)
\]

The following matrices are the transition matrices for Politeness, Sentiment and Emotion for Wicket.
\[ \mathbf{P_{Wicket}} =\left(
\begin{array}{ccc}
0.34	 & 0.3269 &	0.3331\\
0.3089 &	0.3454 &	0.3457\\
0.2889 &	0.3478 &	0.3633
\end{array} \right)
\mathbf{S_{Wicket}} = \left( \begin{array}{lcr}
0.2638 &	0.5569 &	0.1793\\
0.2208 &	0.5795 &	0.1997\\
0.1984 &	0.5393 &	0.2623
\end{array} \right)
\]
\[
\mathbf{E_{Wicket}} = \left( \begin{array}{ccccc}
0.2683 & 0.0113 & 0.0416 & 0.0664 & 0.6123\\
0.1885 & 0.0246 & 0.0656 & 0.0574 & 0.6639\\
0.2299 & 0.0125 & 0.0588 & 0.0481 & 0.6506\\
0.2096 & 0.0092 & 0.0607 & 0.1103 & 0.6103\\
0.1994 & 0.0096 & 0.0523 & 0.0664 & 0.6722
\end{array} \right)
\]

\bibliographystyle{abbrv}

\end{document}